\documentclass[11pt,a4paper]{article}

\usepackage{latexsym}
\usepackage[dvips]{color}
\usepackage{graphicx}

\begin{document}

\title{Semi-flexible hydrogen-bonded and non-hydrogen bonded lattice polymers}

\author{J.\ Krawczyk$^1$, A. L. Owczarek$^1$ and T. Prellberg$^2$\thanks{{\tt {\rm email:}
j.krawczyk@ms.unimelb.edu.au,aleks@ms.unimelb.edu.au,t.prellberg@qmul.ac.uk}} \\
         $^1$Department of Mathematics and Statistics,\\
         The University of Melbourne,\\
         Parkville, Victoria 3052, Australia.\\
$^2$School of Mathematical Sciences\\
Queen Mary, University of London\\
Mile End Road, London E1 4NS, UK
}

\maketitle

\begin{abstract}
  We investigate the addition of stiffness to the lattice model of
  hydrogen-bonded polymers in two and three dimensions.  We find that,
  in contrast to polymers that interact via a homogeneous short-range
  interaction, the collapse transition is unchanged by any amount of
  stiffness: this supports the physical argument that hydrogen bonding
  already introduces an effective stiffness. Contrary to possible
  physical arguments, favouring bends in the polymer does not return
  the model's behaviour to that comparable to the semi-flexible
  homogeneous interaction model, where the canonical $\theta$-point
  occurs for a range of parameter values. In fact, for sufficiently
  large bending energies the crystal phase disappears altogether, and
  no phase transition of any type occurs. We also compare the
  order-disorder transition from the globule phase to crystalline
  phase in the semi-flexible homogeneous interaction model to that for
  the fully-flexible hybrid model with both hydrogen and non-hydrogen
  like interactions. We show that these phase transitions are of the
  same type and are a novel polymer critical phenomena in two
  dimensions. That is, it is confirmed that in two dimensions this
  transition is second-order, unlike in three dimensions where it is
  known to be first order.  We also estimate the crossover exponent
  and show that there is a divergent specific heat, finding
  $\phi=0.7(1)$ or equivalently $\alpha=0.6(2)$. This is therefore
  different from the $\theta$ transition, for which $\alpha=-1/3$.

\end{abstract}

\newpage

\section{Introduction}

When modelling a single polymer in dilute solution by a lattice
self-avoiding walk \cite{Flory:book,deGennes:book,Vanderzande:book}
the effects of solvent mediated intra-polymer interaction are often
included by assigning an energy to each non-consecutive pair of
monomers lying on the neighbouring lattice sites: the interaction is
homogeneous in that in doesn't depend on shape of the two parts of the
polymer containing the interacting monomers. This is the well-studied
\emph{Interacting self-avoiding walk (\textit{ISAW}) model} which is
the standard model of polymer collapse using self-avoiding walks. If
the energy is repulsive the polymer behaves as a swollen chain (the
so-called excluded-volume state) regardless of temperature and one
says that it is in a good solvent. When the energy is attractive, and
the temperature is low enough, the chain becomes a rather more compact
globule \cite{deGennes:book,desCloizeaux:book}, reminiscent of a
liquid droplet: this is also known as the poor solvent situation. The
transition point between these two phases is called the
$\theta$-point; it is a well studied continuous phase transition (see
\cite{prellberg:1994-01} and references therein).  However, the
mapping of monomers onto lattice sites ignores the natural rigidity of
real polymers. An energy for bends in the self-avoiding walk can be
introduced to take account of this feature.  The model of
semi-flexible polymers has been investigated mostly in three
dimensions \cite{bastolla:1997-01,vogel_2007:ss,doye_1997:ss}. In
particular, Bastolla and Grassberger \cite{bastolla:1997-01}
investigated semi-flexible interacting self-avoiding walks
(semiflexible \textit{ISAW}) on the cubic lattice, which interact via
all nearest-neighbours, as in the $\theta$-point model, and included
the bending energy. They showed that when there is a strong energetic
preference for straight segments, this model undergoes a single
first-order transition from the excluded-volume high-temperature state
to a crystalline state.  Intriguingly, if there is only a weak
preference for straight segments, the polymer undergoes two phase
transitions: on lowering the temperature the polymers undergoes the
$\theta$-point transition to the liquid globule followed at a lower
temperature by a first-order transition to the frozen crystalline
phase. In two dimensions the transition between the globule and the
frozen state has been studied in Hamiltonian walks, and there it seems
to be a continuous one \cite{jacobsen:2004-01}.

The modelling of polymers in solution changes as soon as we want to
describe any biological system (\textit{e.g.\ }proteins), in which the
hydrogen bonding plays an important role \cite{pauling:1951-01}. One
of the main features of the bonding is that the interacting residua
lie on a partially straight segments of the chain.  Hydrogen-like
bonding was first modelled on the cubic and square lattices using
Hamiltonian paths by Bascle \textit{et al.\ }\cite{bascle:1993-01}.  A
monomer acquires a hydrogen-like bond with its (non-consecutive)
nearest neighbour if both of them lie on straight sections of the
chain.  Note that the identification of a single contact of this type
with a single hydrogen bond is only valid if fully-flexible polymers
are considered; otherwise the contact represents an agglomeration of
such bonds. The interacting self-avoiding walk modified to have only
such interactions will be referred to as the hydrogen-like bonding
model, or rather \textit{IHB} model.  The \textit{IHB} model was
studied in mean-field approximation \cite{bascle:1993-01} and a
first-order transition from a high-temperature excluded-volume
(swollen) phase to a quasi-frozen solid-like phase was found in both
two and three dimensions.  Hence this would indicate that it is a
different type of transition from the $\theta$-point. Note also that
the low temperature phase was found to be anisotropic whereas the
collapsed globule of the standard $\theta$-point model is isotropic.
The \textit{IHB} model on the square lattice was studied directly by
Foster and Seno by means of the transfer matrix method
\cite{foster:2001-01} and by Krawczyk \textit{et al.\ 
}\cite{my:2007-01} on both the square and cubic lattice using a Monte
Carlo method. In both of these studies a first-order transition was
found between an excluded-volume (swollen-coil) state and an
anisotropic ordered compact phase in two and in three dimensions,
again in opposition to the $\theta$-point \cite{deGennes:book}.

However, the \textit{IHB} model was recently extended
\cite{krawczyk_2007:hb} to a hybrid model (\textit{IHB--INH}) that
includes both the hydrogen-like bond interactions and non-hydrogen
like bond interactions, with separate energy parameters. When the
non-hydrogen bonding energy is set to zero the \textit{IHB} model is
recovered. If both energies are set to be the same then the
\textit{ISAW} \emph{without stiffness} is recovered. For large values
of the ratio of the interaction strength of hydrogen-bonds to
non-hydrogen bonds, a polymer will undergo a single first-order phase
transition from a swollen coil at high temperatures to a folded
crystalline state at low temperatures.  On the other hand, for any
ratio of these interaction energies less than or equal to one there is
a single $\theta$-like transition from a swollen coil to a liquid
droplet-like globular phase. Importantly, for intermediate ratios two
transitions can occur, so that the polymer first undergoes a
$\theta$-like transition on lowering the temperature, followed by a
second transition to the crystalline state. In three dimensions it was
found that this second transition is first order, while in two
dimensions it is probably second order with a divergent specific heat.
In other words, at least in three dimensions, by adding an energy to
both the hydrogen-like and non-hydrogen like interactions a phase
diagram similar to the one for the semi-flexible \textit{ISAW} is
obtained.

Various issues then arise. Firstly it is worth studying the addition
of stiffness to the \textit{IHB} model since stiffness clearly affects
the type of phases that occur in the \textit{ISAW} model. Moreover,
one could argue that since in the \textit{IHB} model interactions
occur only between straight segments of the walk, an effective
stiffness has already been introduced: one could then go further and
argue that favouring bends may result in behaviour like that in the
\textit{ISAW} model. Secondly, it is worth investigating the
semi-flexible \textit{ISAW} model in two dimensions to check if the
similarity of the \textit{IHB--INH} phase diagram to the semi-flexible
\textit{ISAW} extends to that dimension. In particular, given the
existence of the same three phases (as we shall find) whether the
globule-crystal transition is of the same second order type as in the
\textit{IHB--INH} model. Finally the more general relationship between
these three potentials (nearest-neighbour interaction, hydrogen-like
bonding and stiffness) on the phase structure of the model polymers is
worth pursuing.  Hence, in this paper we investigate via Monte Carlo
simulations the effect of adding stiffness to the interacting hydrogen
bonding model and compare this to adding stiffness to the canonical
\textit{ISAW} model. Intriguingly we find that favouring stiffness
does \emph{not change} the single transition found in the \textit{IHB}
model without stiffness.  However, for a sufficiently large bending
energy favouring bends the transition temperature is found to go to
zero, and for larger ratios of bending energy to hydrogen bonding
energy \emph{no} phase transition occurs. We then compare the
globule-crystal phase transition in the semi-flexible \textit{ISAW}
model on the square lattice (on the cubic lattice they are both
first-order) to that of the \textit{IHB--INH} model and show that the
exponents are most likely the same: we also estimate these exponents.

In general we have considered various restrictions of a model of three
parameters where stiffness is added to the hydrogen-bond interactions
and non-hydrogen-bond interactions.  In our conclusions we argue that
only three phases exist in the larger parameter space: an
excluded-volume dominated state (where the polymer is `swollen'), a
disordered globular state where the polymer is in a condensed
liquid-like drop and a crystalline state.

The paper is organised as follows. In Section 2 we explain more
carefully details of the models considered.  In Section 3 we consider
the semi-flexible interacting hydrogen bond model. In Section 4 we
consider the semi-flexible version of the canonical interacting SAW
model and we compare our results to those from a model that has
different energies for hydrogen and non-hydrogen nearest-neighbour
interactions. In Section 5 we investigate a semi-flexible interacting
polymer where there are no hydrogen bonds. We end with some
conclusions about the most general model defined in Section 2.

\section{Definitions}
Let us define a general model that contains each of the models considered as sub-cases via restricting parameters. 

\begin{figure}[ht!]
  \centering
  \includegraphics[scale=1.0]{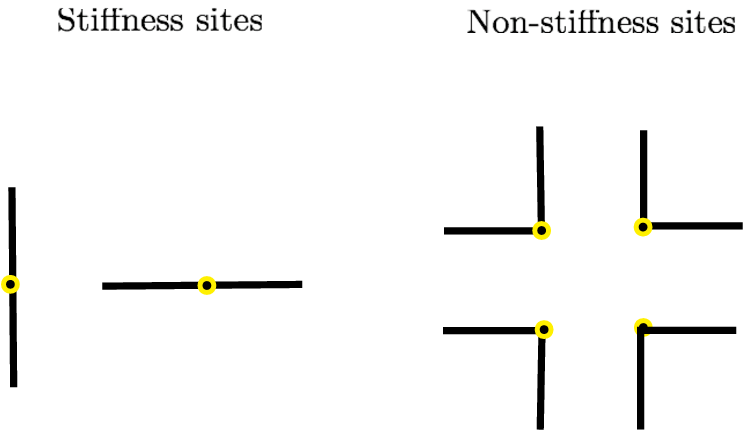}
  \caption{Stiffness sites and non stiffness sites on the square
    lattice} 
  \label{stiffness_sites}
\end{figure}

We begin with a self-avoiding walk on the square and simple cubic
lattices. The walk consists of a sequence of occupied lattice sites
joined by \emph{steps} of the walk. A walk of $n$ steps occupies $n+1$
sites. Consider the sites of the lattice occupied by the walk. When
two sites of the walk are adjacent on the lattice and not consecutive
along the walk, so as not to be joined by a step of the walk, we
refer to this pair of sites as a nearest-neighbour \emph{contact}.
Additionally, let us refer to two consecutive steps that follow the
same lattice direction as a \emph{stiff step-pair}, so that there are
three consecutive occupied sites along a line on the lattice, and the
site in the centre of this trio to be a \emph{stiffness} site (see
Figure~\ref{stiffness_sites}).

\begin{figure}[ht!]
  \centering
  \includegraphics[scale=1.0]{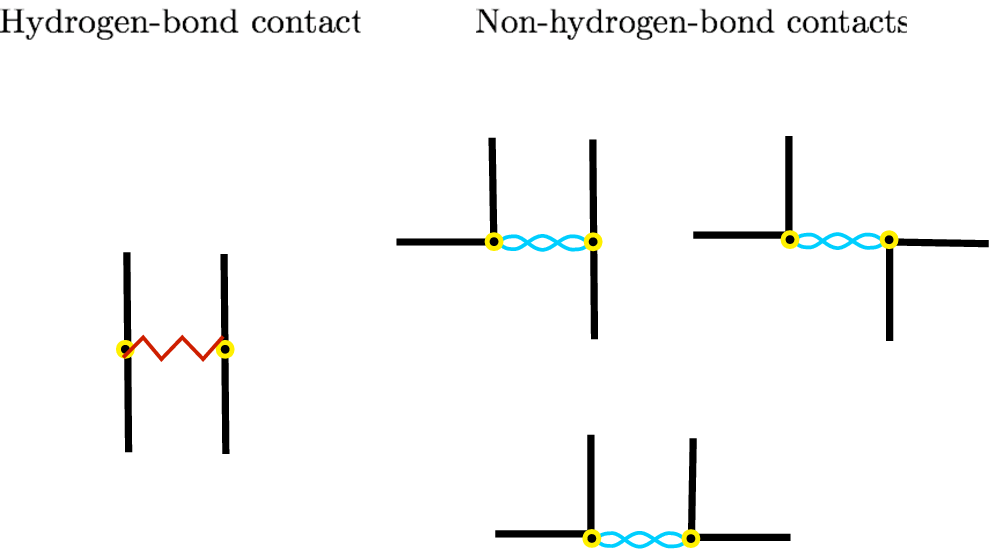}
  \caption{The partition of the types of nearest-neighbour contacts
    into hydrogen-bonds and non-hydrogen-bonds  for the square lattice
    is shown. Rotations of these are also possible.} 
  \label{contact_types}
\end{figure}

\begin{figure}[ht!]
  \centering
  \includegraphics[scale=0.5]{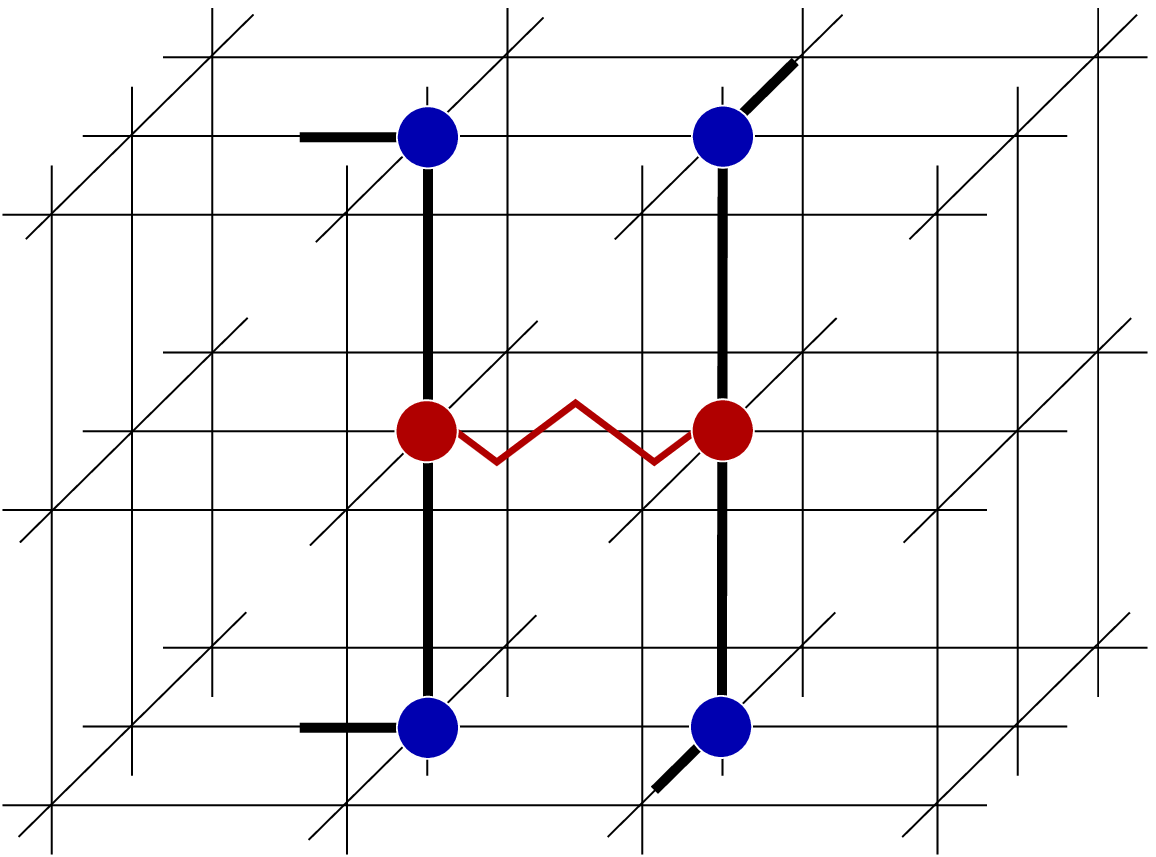}\hspace*{0.3cm}
  \includegraphics[scale=0.5]{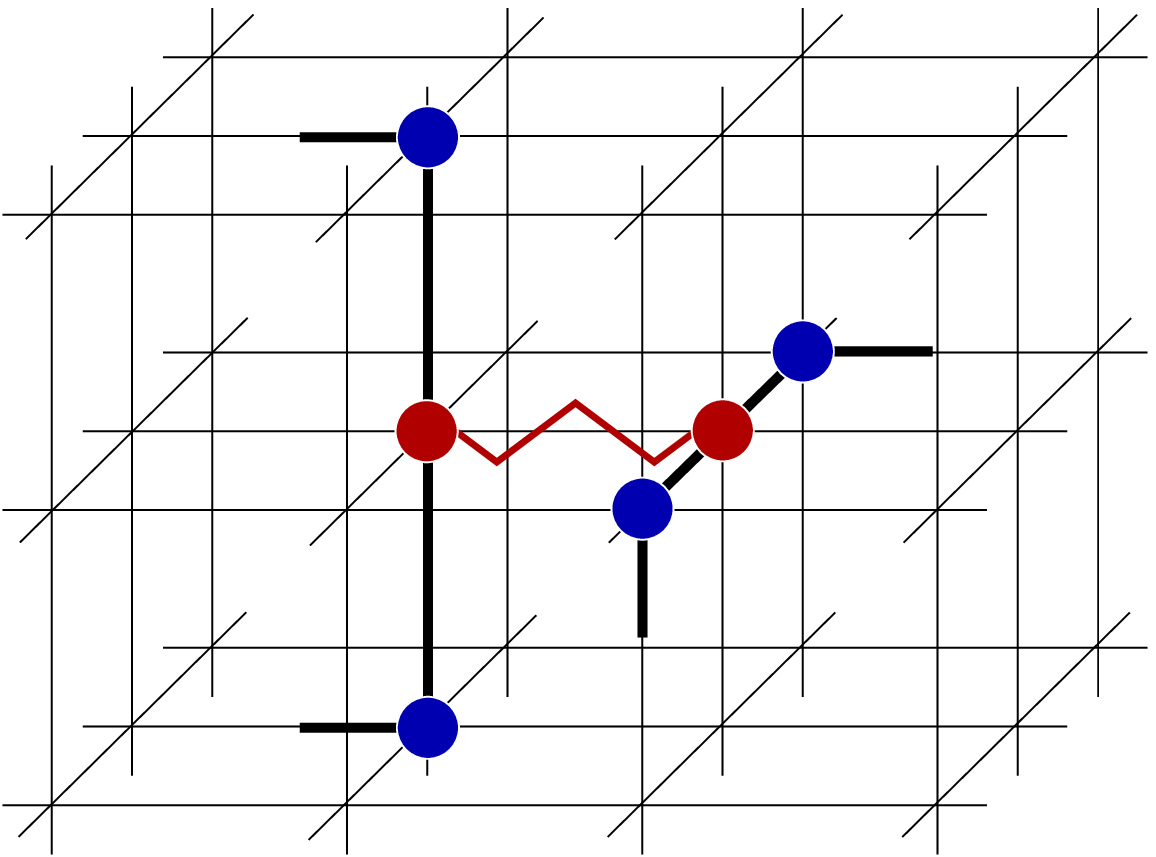}
  \caption{Hydrogen-bond contacts for the simple cubic lattices.}
  \label{3dcontact_types}
\end{figure}

Now partition the possible types of contact into two classes: when
they occur between stiffness sites then we refer to these as
\emph{hydrogen-bond} contacts, and all others are
\emph{non-hydrogen-bond} contacts. In Figure~\ref{contact_types} the
partition of the types of nearest-neighbour contacts for the square
lattice is shown. In Figure~\ref{3dcontact_types} the hydrogen-bond
contacts (only) are shown for the cubic lattice.

\begin{figure}[ht!]
  \centering
  \includegraphics[scale=1.0]{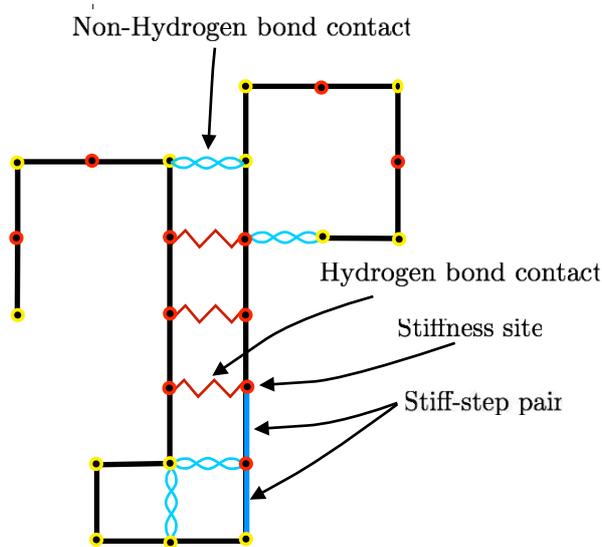}
  \caption{A sample walk configuration with the various type of
    interaction highlighted for the square lattice.} 
  \label{sample_walk}
\end{figure}

We now add an energy to the self-avoiding which consists of three
contributions: an energy for each hydrogen-bond contact
$-\varepsilon_{hb}$, an energy of non-hydrogen-bond contact
$-\varepsilon_{nh}$ and an energy for stiff step-pairs
$-\varepsilon_{ss}$. The total energy of a walk configuration
$\varphi_n$ of $n$ steps is
\begin{equation}
E_n(\varphi_n) = -m_{hb}(\varphi_n)\ \varepsilon_{hb}
-m_{nh}(\varphi_n)\ \varepsilon_{nh}- m_{ss}(\varphi_n)\
\varepsilon_{ss}, 
\end{equation}
where $m_{hb}$ denotes the number of of hydrogen-bond contacts,
$m_{nh}$ denotes the the number of non-hydrogen-bond contacts, and
$m_{ss}$ denotes the number of stiffness sites in the walk
configuration~(see~Figure~\ref{sample_walk}).

We will denote the total number of all nearest-neighbours interactions
as $m_{is}$, where it is equal to the sum of the number of the two
types of interaction considered in our full model, that is
$m_{is}=m_{nh}+m_{hb}$.

The inverse temperature is denoted as $\beta=1/k_{B}T$, where $k_B$ is
the Boltzmann constant and $T$ the absolute temperature.  We define
for convenience $\beta_{hb}=\beta\varepsilon_{hb}$,
$\beta_{nh}=\beta\varepsilon_{nh}$ and
$\beta_{ss}=\beta\varepsilon_{ss}$.  The partition function is then
given by
\begin{equation}
Z_n(\beta_{hb},\beta_{nh},\beta_{ss})=
\sum_{m_{hb},m_{nh},m_{s}} C_{n,m_{hb},m_{nh},m_{ss}}\; e^{\beta_{hb}
  m_{hb}+\beta_{nh} m_{nh}+\beta_{ss}m_{ss}} 
\label{part-func}
\end{equation}
with $C_{n,m_{hb},m_{nh},m_{ss}}$ the density of states. Canonical averages
are calculated with respect to this density of states.

Our results are for the following models:
\begin{itemize}
\item the semi-flexible interacting hydrogen-bonding model
(semi-flexible
 \textit{IHB} model) where $\beta_{nh}=0$,
\item the semi-flexible interacting non-hydrogen-bonding model
  (semi-flexible \textit{INH} model) where $\beta_{hb}=0$,
\item and the semi-flexible interacting self-avoiding walk model
  (semi-flexible \textit{ISAW} model) where $\beta_{hb}=\beta_{nh}$.
\end{itemize}
We remind the reader that the semi-flexible \textit{ISAW} model on the
simple cubic lattice has previously been studied by Bastolla and
Grassberger \cite{bastolla:1997-01}.

We compare our results to those concerning the previously studied
\cite{krawczyk_2007:hb} model where different energies are assigned to
hydrogen-bond contacts and non-hydrogen-bond contacts, though not to
stiffness sites, namely
\begin{itemize}
\item the Interacting hydrogen-bonding -- Interacting
  non-hydrogen-bonding model (\textit{IHB--INH} model) where
  $\beta_{ss}=0$. 
\end{itemize}

All the simulations in this paper use a Monte Carlo technique, known
as FlatPERM \cite{my:2004-01}, which is well suited to the study of
self-avoiding walks on the simple cubic and square lattices with
interactions. This technique allows for the estimation of quantities
at all values of an interaction parameter by the estimation of the
appropriate `density of states': \textit{e.g.}\ from equation
(\ref{part-func}) the estimation of $C_{n,m_{hb},m_{nh},m_{ss}}$ would
allow the calculation of canonical averages for any value of the
parameters $\beta_{hb}$, $\beta_{nh}$ and $\beta_{ss}$. On the other
hand each new parameter increases the computational cost by at least a
factor of $n$ (being the range of the variable --- \textit{e.g.}\ 
$m_{hb}$ when including $\beta_{hb}$). Therefore so as to obtain data
for reasonable length walks it is necessary to restrict simulations to
one or two interaction parameters.


\section{Semi-flexible \textit{IHB} model ($\beta_{nh}=0$)}
We have simulated the semi-flexible \textit{IHB} model on the square
and simple cubic lattice, estimating an appropriate density of states
$C_{n,m_{hb}, m_{ss}}$, so that averages can be performed for all
values of the parameters $\beta_{hb}$ and $\beta_{ss}$. We have
estimated $\langle m_{hb} \rangle$ and $\langle m_{ss} \rangle$, which
are directly related to the internal energy, and the variances in
these averages which are related to the specific heat of the model.
These `two parameter' simulations were completed up to $n=128$ steps.

\begin{figure}[ht!]
  \centering
  \includegraphics[scale=0.6]{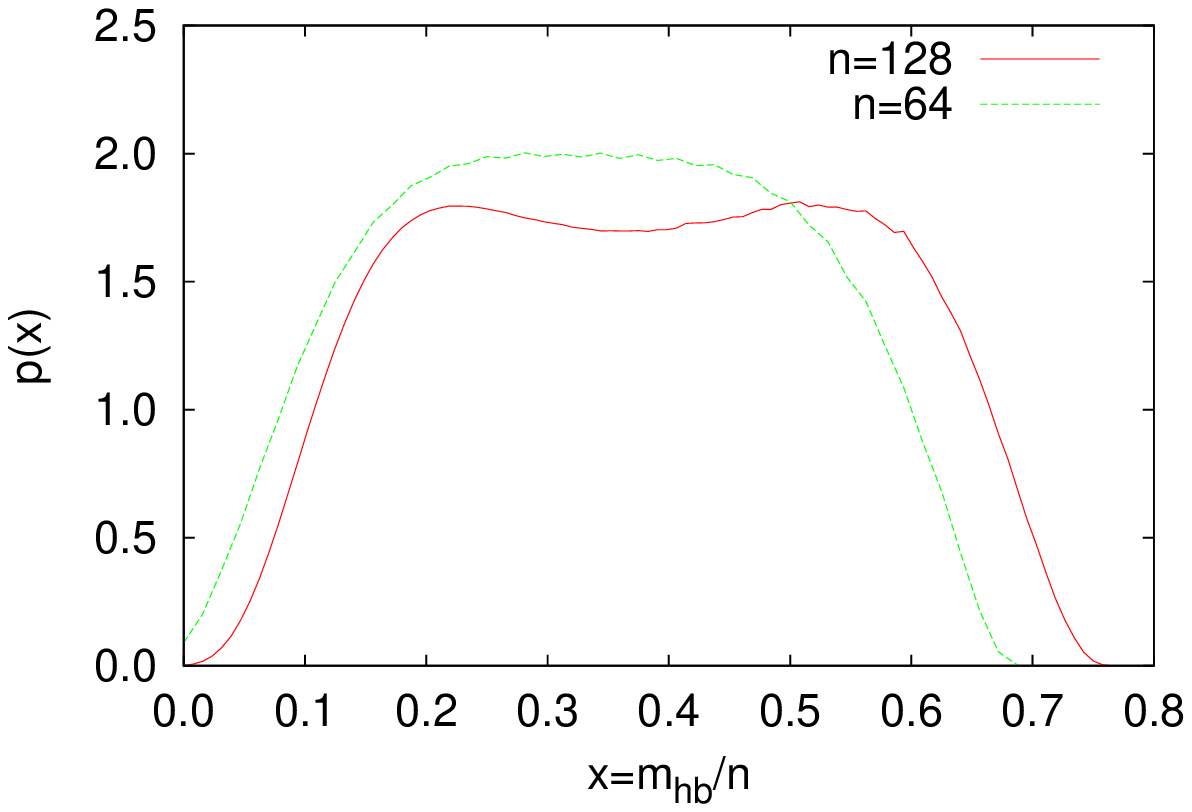}
  \includegraphics[scale=0.6]{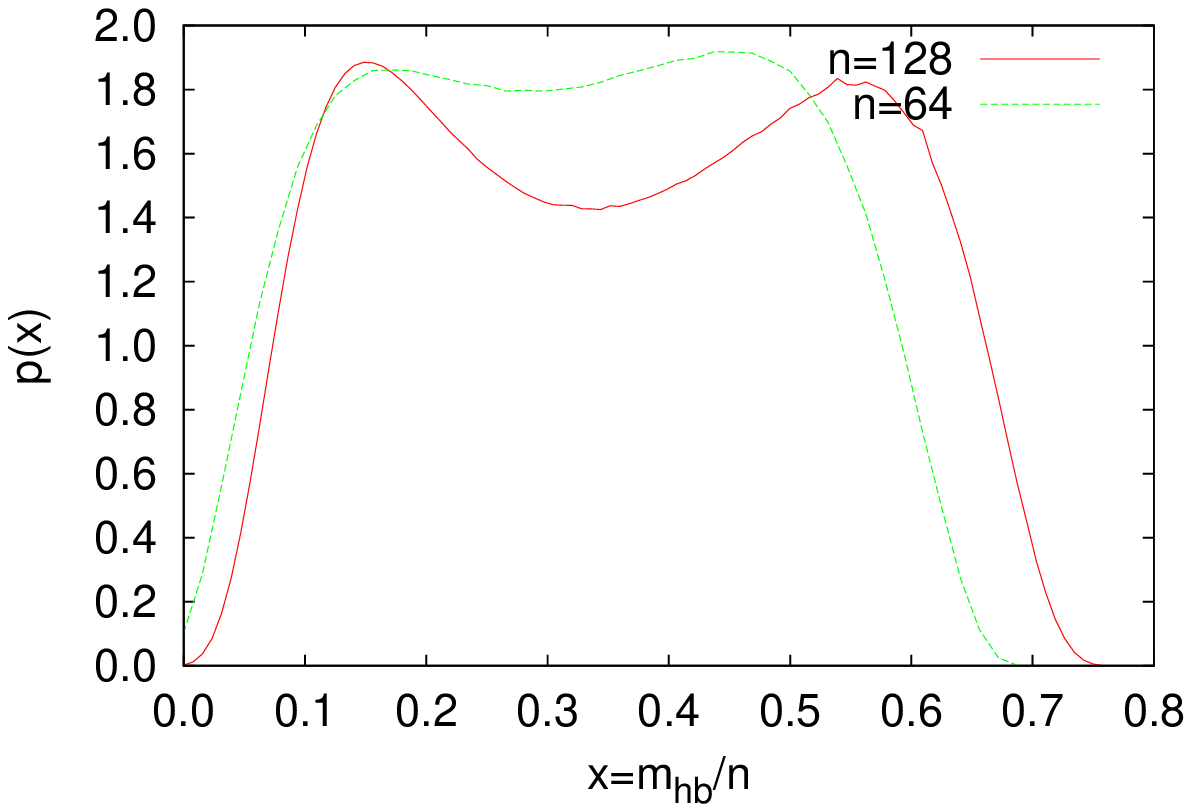}
  \caption{Distribution of internal energy for the square lattice
    semi-flexible \textit{IHB} model at the estimated transition point
    $\beta_{hb}^{(c)}(\beta_{ss}; n)$ when  $\beta_{ss}=0.5$ (left),
    and $\beta_{ss}=-0.5$ (right), using $n=128$ and $n=64$.} 
  \label{hb_ss_hist}
\end{figure}

Let us start by considering the square lattice data. As has been
previously found \cite{foster:2001-01,my:2007-01} at $\beta_{ss}=0$,
we find that for any fixed value of $\beta_{ss}$ there is strongly
growing peak in the variance of $m_{hb}$ at a single value of
$\beta_{hb}$. This is indicative of a phase transition at some
position $\beta_{hb}^{(c)}(\beta_{ss})$ and as the normalised peak of
the variance is growing close to linearly with $n$ irrespective of
$\beta_{ss}$ it is indicative of a first order phase transition. To
confirm this we considered the distribution of $m_{hb}$ at the
finite-size transition point $\beta_{hb}^{(c)}(\beta_{ss}; n)$. In
Figure~\ref{hb_ss_hist} we show this distribution at
$\beta_{hb}^{(c)}(0.5; 128)$, $\beta_{hb}^{(c)}(0.5; 64)$ and
$\beta_{hb}^{(c)}(-0.5; 128)$, $\beta_{hb}^{(c)}(-0.5; 64)$.  The
deepening bimodal distributions reinforce our conclusion that the
transition is first order regardless of $\beta_{ss}$.

By considering the scaling of the end-to-end distance and the
anisotropy parameter (see \cite{krawczyk_2007:hb}) we have verified
that for $\beta_{hb} < < \beta_{hb}^{(c)}(\beta_{ss},n)$ the extended
phase exists while for $\beta_{hb} >> \beta_{hb}^{(c)}(\beta_{ss},n)$
the anisotropic crystal phase is observed.

\begin{figure}[ht!]
  \centering
   \includegraphics[scale=0.7]{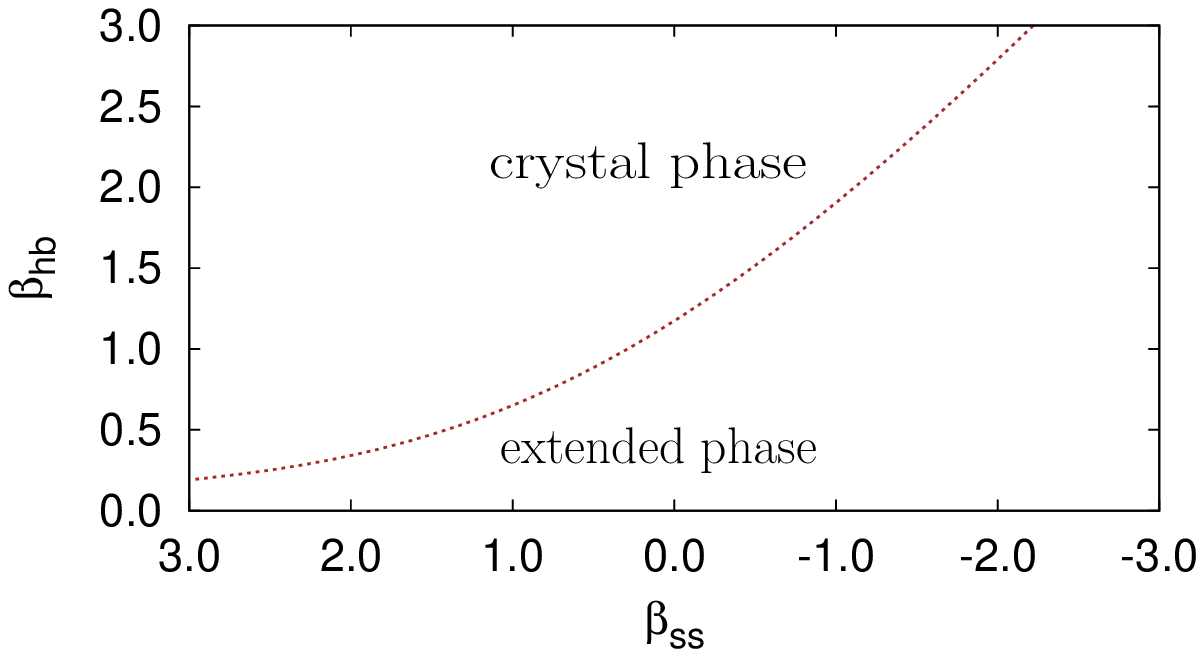}
    \includegraphics[scale=0.3]{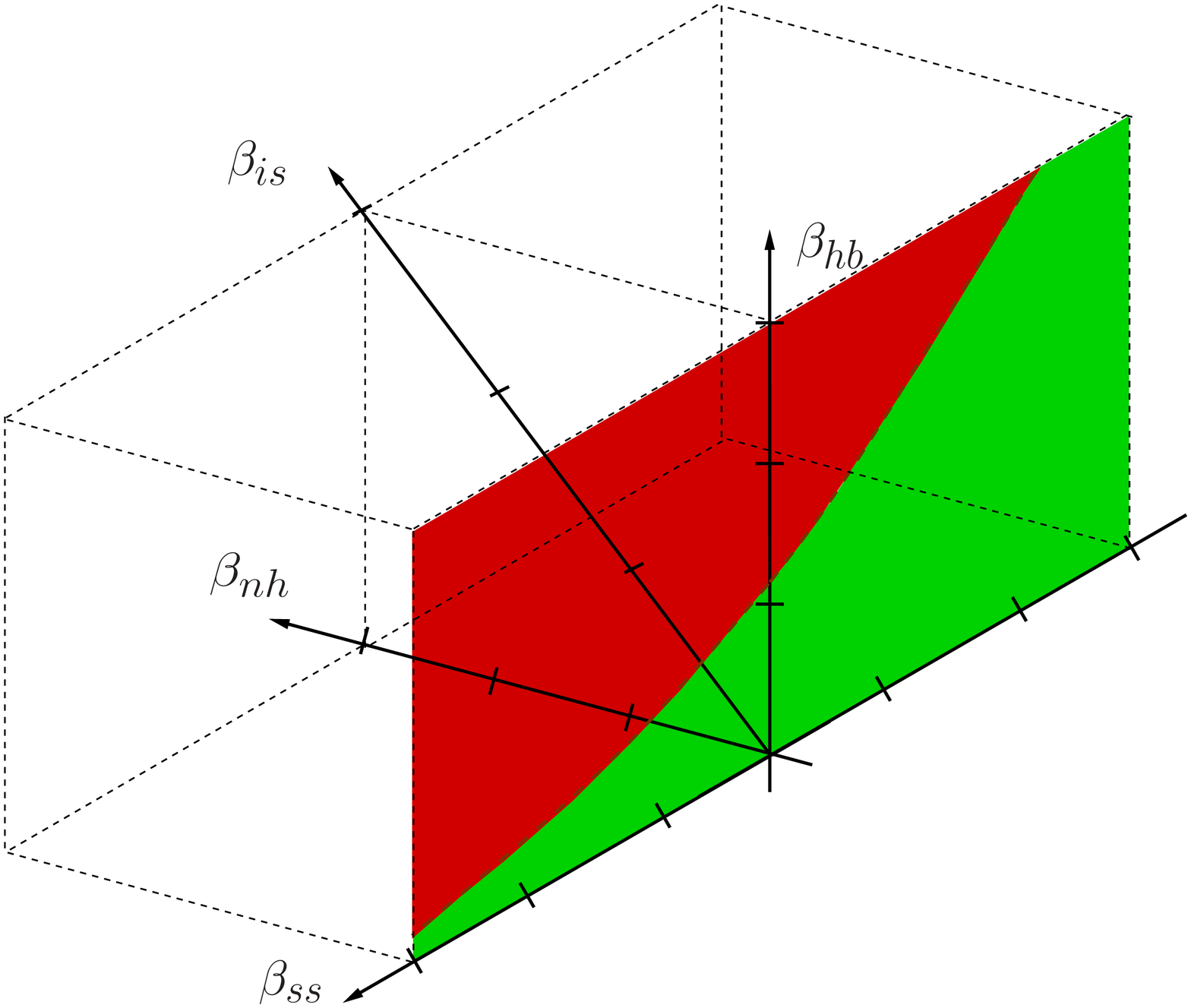}
  \caption{
  Plot of the finite size phase boundary for the semi-flexible
  \textit{IHB} model on the square lattice (left). The plane of the
  parameters of the semi-flexible \textit{IHB} model in the more
  general three parameter space: the two phases are denoted by a dark
  (red) shading for the crystalline phase and a mid-density (green)
  shading for the extended swollen polymer phase. (right) 
 } 
  \label{fig_mod_sshb}
\end{figure}

In Figure~\ref{fig_mod_sshb} a plot of $\beta_{hb}^{(c)}(\beta_{ss};
128)$ is given in the two-dimensional space of $\beta_{hb}$, and
$\beta_{ss}$. Next to that in Figure~\ref{fig_mod_sshb} the same curve
is plotted in the three-dimensional space of $\beta_{hb}$,
$\beta_{nh}$ and $\beta_{ss}$.

\begin{figure}[ht!]
  \centering
  \includegraphics[scale=0.6]{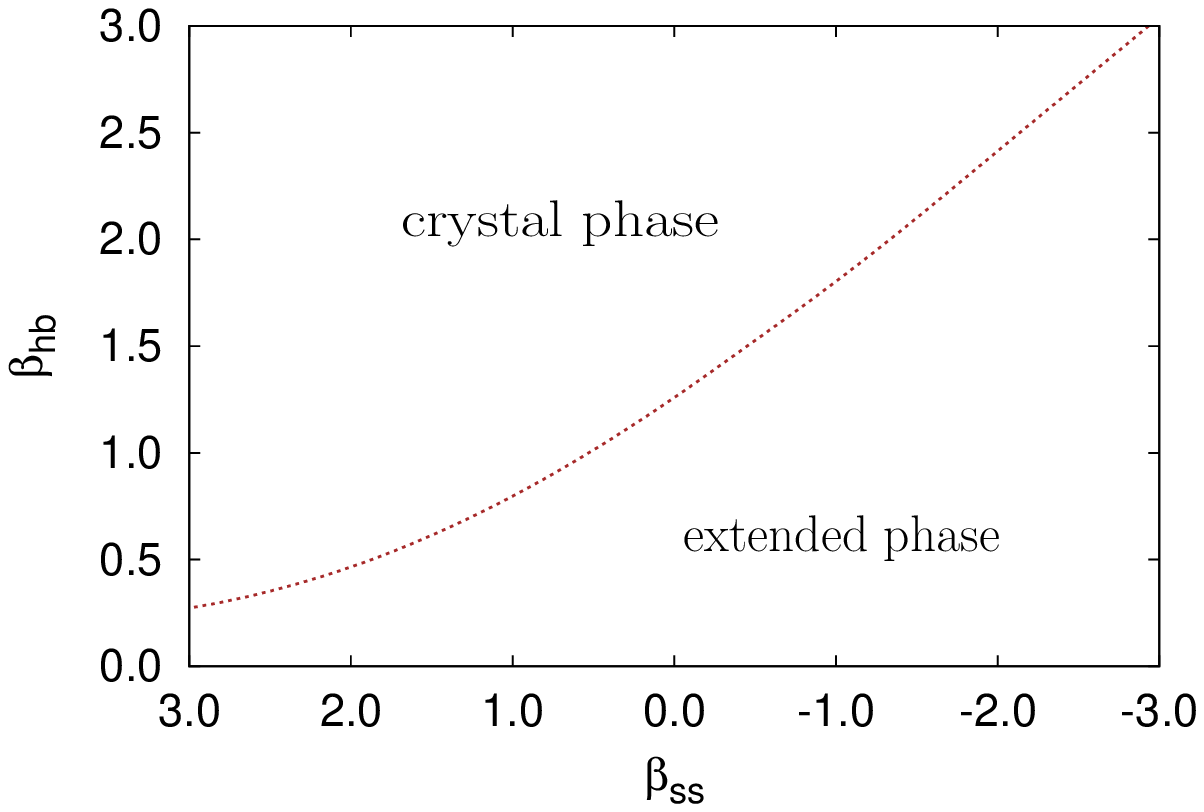}
  \caption{Plot of the finite size phase boundary for the
    semi-flexible \textit{IHB} model on the cubic lattice}
   \label{fig_mod_3sshb}
\end{figure}

The results for the cubic lattice are completely analogous and the
finite size phase diagram is given in Figure~\ref{fig_mod_3sshb}.  

On both lattices we therefore have found that the addition of positive
stiffness to the interacting hydrogen-like bond model leaves the
single phase transition from a swollen phase at high temperatures to a
crystalline phase at low temperatures unchanged. When favouring bends,
so that $\varepsilon_{ss}$ and $\beta_{ss}$ are negative, then when
any transition occurs it is again of a similar type as the
fully-flexible model. However there is a range of $\lambda =
\varepsilon_{ss}/\varepsilon_{hb}$ such that no transition occurs.
This is relative easy to understand: hydrogen bonds only occur between
two stiffness sites which are suppressed by large negative values of
$\varepsilon_{ss}$. Let us focus on the square lattice. On the square
lattice for $\lambda = \varepsilon_{ss}/\varepsilon_{hb}< -1 $ no
phase transition occurs on lowering the temperature.  At zero
temperatures the ground state for $\lambda<-1$ is a walk consisting
only of bends with energy zero (this state should have a positive
entropy). For $-1< \lambda$ the ground state is one with long folds
($\beta$-like sheets) and a bulk energy $-n
(\varepsilon_{hb}+\varepsilon_{ss})$ which is negative so long as
$\varepsilon_{hb}> -\varepsilon_{ss}$. This state has zero entropy.
The difference in entropy accounts for the apparent shift of the
asymptote of the phase boundary from $\beta_{hb}= -\beta_{ss}$ to
$\beta_{hb}= -\beta_{ss}+c$ in Figure~\ref{fig_mod_sshb}.


\section{Semi-Flexible \textit{ISAW} ($\beta_{nh}=\beta_{hb}$)}

For the semi-flexible \textit{ISAW} model we focus our attention on
the square lattice which has not previously been investigated. On the
square lattice we performed simulations for $n=128$ for two parameters
$\beta_{is}$ and $\beta_{ss}$ as well as for the one parameter
$\beta_{ss}$ for constant $\beta_{is}=0.7$ for lengths up to $n=512$.
The line $\beta_{is}=0.7$ was chosen so as to focus on the transition
between two collapsed phases: the collapsed-globule and the crystal.

\begin{figure}[ht!]
  \centering
  \includegraphics[scale=0.7]{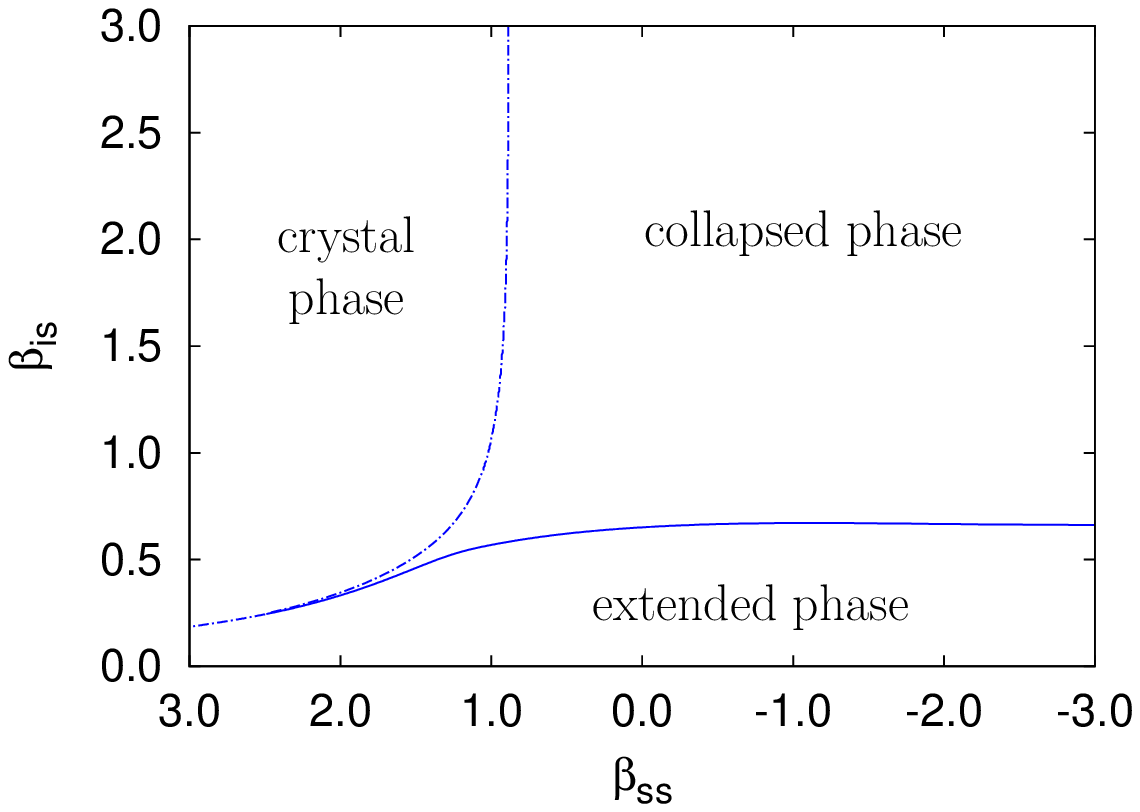}
    \includegraphics[scale=0.3]{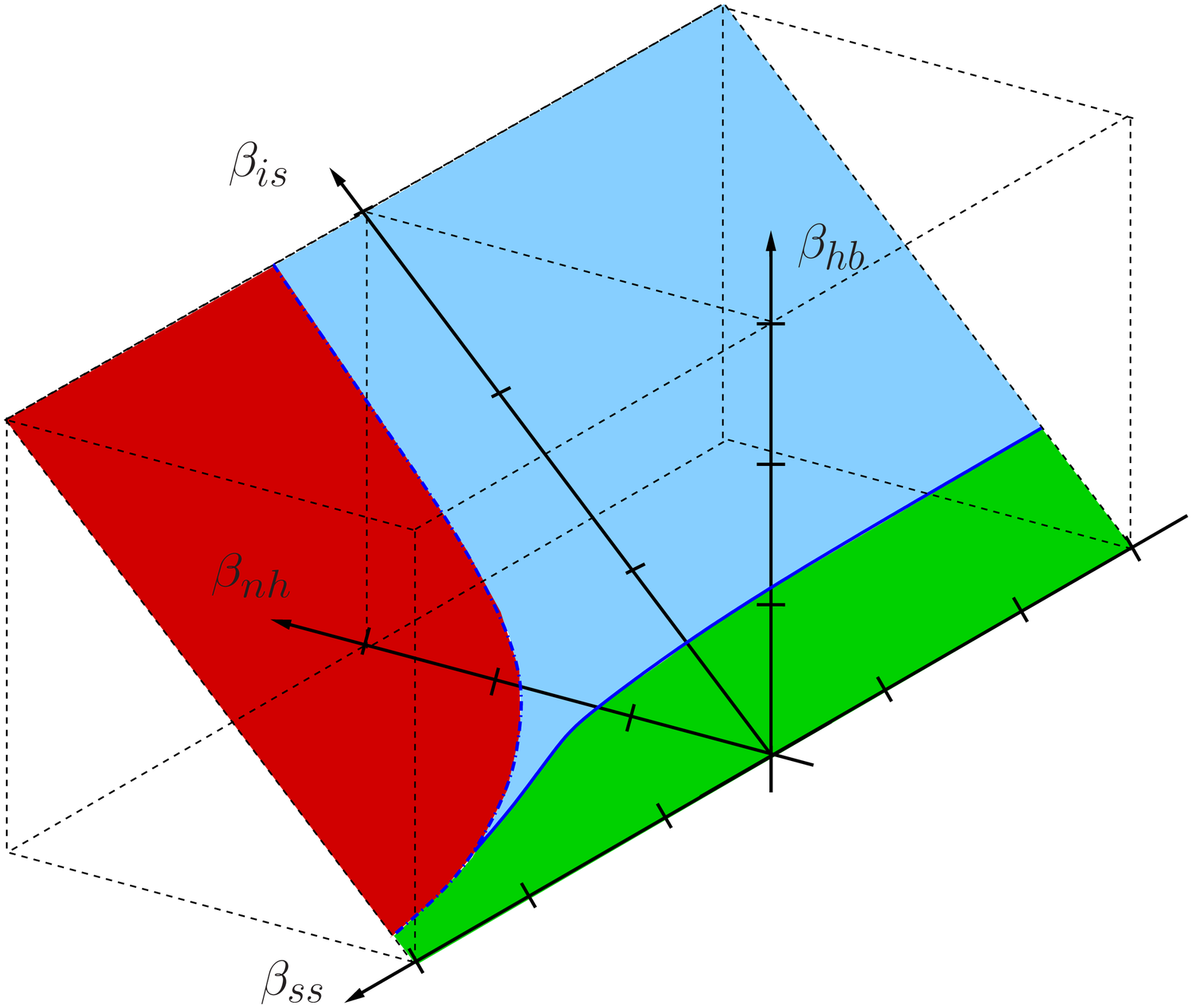}
  \caption{Plot of the finite size phase boundary for the
    semi-flexible \textit{ISAW} model on the square lattice (top). The
    plane of the parameters of the semi-flexible \textit{ISAW} model
    in the more general three parameter space: the three phases are
    denoted by a light (blue) shading for the globular phase, a
    mid-density (green)  shading for the extended swollen polymer
    phase and a dark (red) shading for the crystalline
    phase. (bottom)} 

  \label{fig_mod_isss}
\end{figure}

Let us begin with the two parameter simulations. By considering the
maximum eigenvalue of the matrix of fluctuations of $m_{is}$ and
$m_{ss}$ we have mapped out a finite sized phase boundary, see
Figure~\ref{fig_mod_isss}.

The boundaries clearly divide the phase space into three phases. Once
again, by considering the scaling of various quantities such as the
end-to-end distance at fixed points deep within each suspected phase
we are satisfied that the three phases are the same as in the
\textit{IHB--INH} model \cite{krawczyk_2007:hb} on the square lattice:
a swollen phase with $\nu=3/4$, and two collapsed phases where
$\nu=1/2$. In one phase the typical configurations are clearly
anisotropic, looking like folded $\beta$-sheets, indicating that it is
a crystalline phase. Hence the phase structure and phase diagram is
similar to the three-dimensional case.  We have attempted to locate
the triple point which seems to around $(\beta_{ss},\beta_{is})
=(2.0,0.3)$. However using different methods has resulted in quite
different estimates so we do not propose any error estimate on these values.

\begin{figure}[ht!]
  \centering
  \includegraphics[scale=0.7]{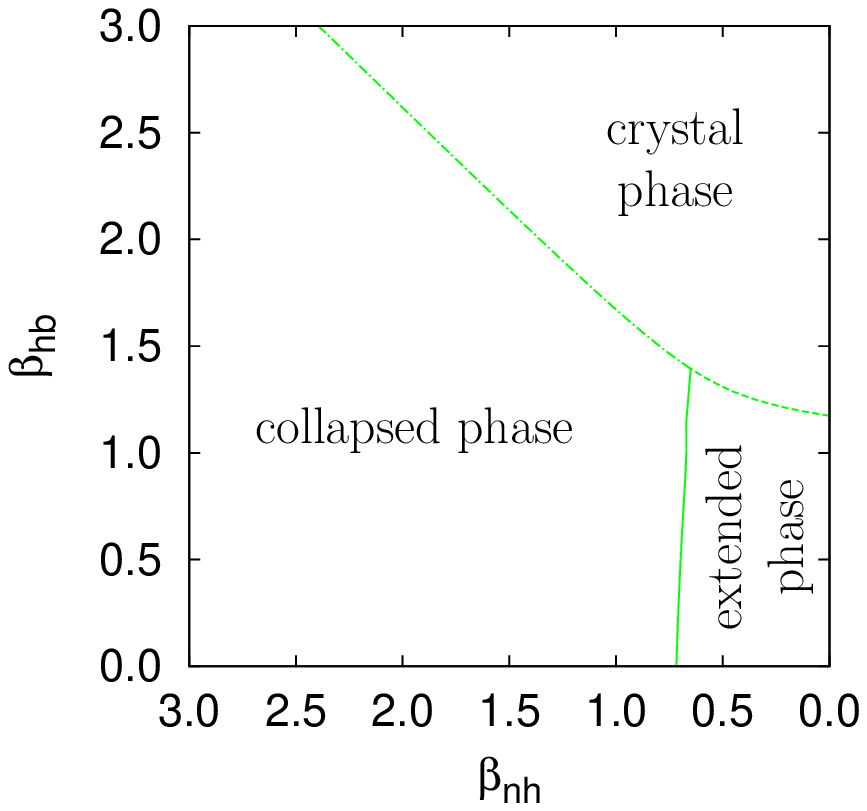}
    \includegraphics[scale=0.3]{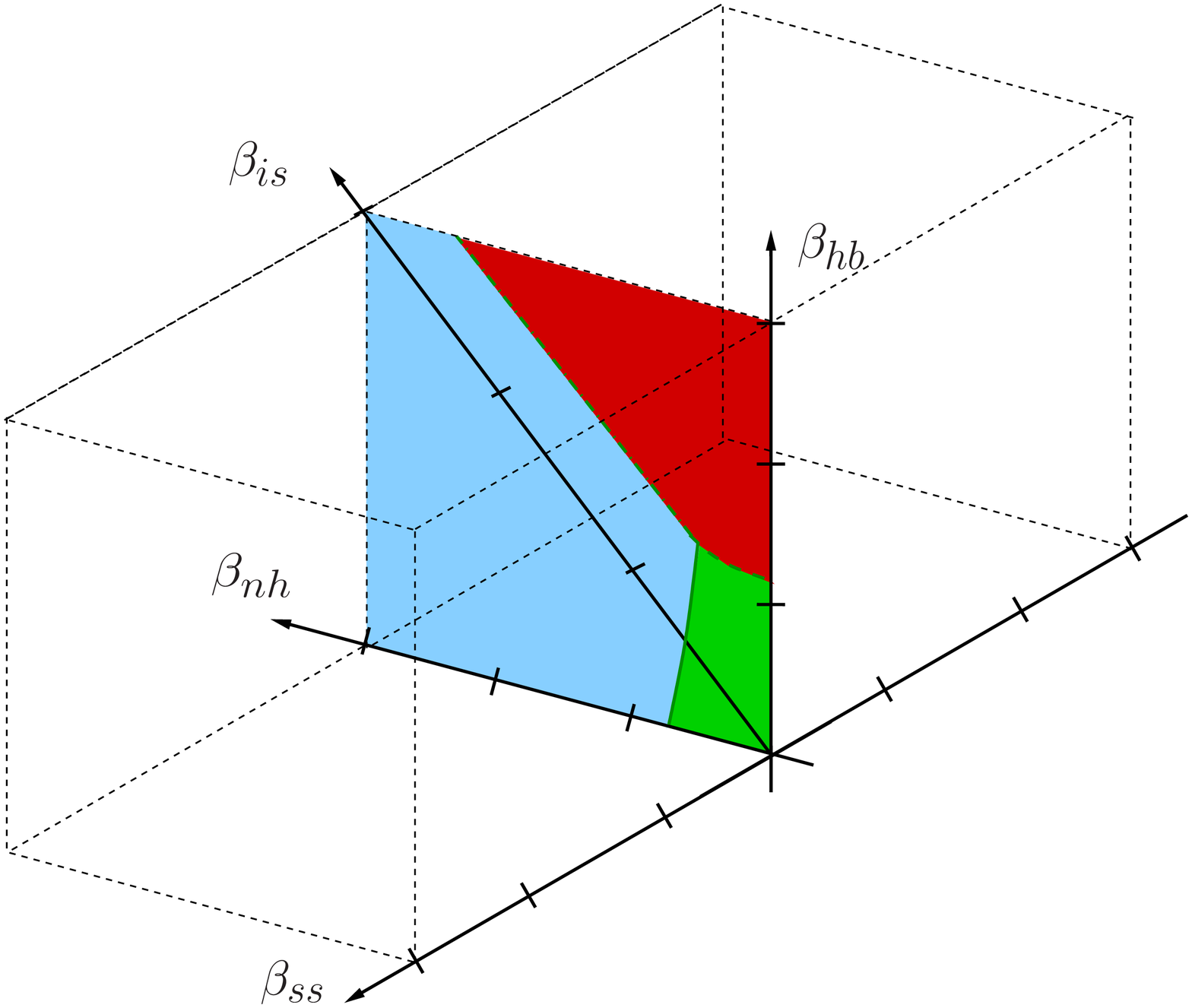}
  \caption{Plot of the finite size phase boundary for the
    fully-flexible \textit{IHB--INH} model on the square lattice
    (top). The plane of the parameters of the semi-flexible
    \textit{INH} model in the more general three parameter space: the
    three phases are denoted by a light (blue) shading for the
    globular phase, a mid-density (green)  shading for the extended
    swollen polymer phase and a dark (red) shading for the crystalline
    phase (bottom).} 
   \label{fig_mod_hbis}
\end{figure}

For convenience and comparison the corresponding finite size phase
boundary diagrams for the \textit{IHB--INH} model studied by Krawczyk
{\it et al.} \cite{krawczyk_2007:hb} are given in
Figure~\ref{fig_mod_hbis} .

\begin{figure}[t!]
  \centering
  \includegraphics[scale=0.8]{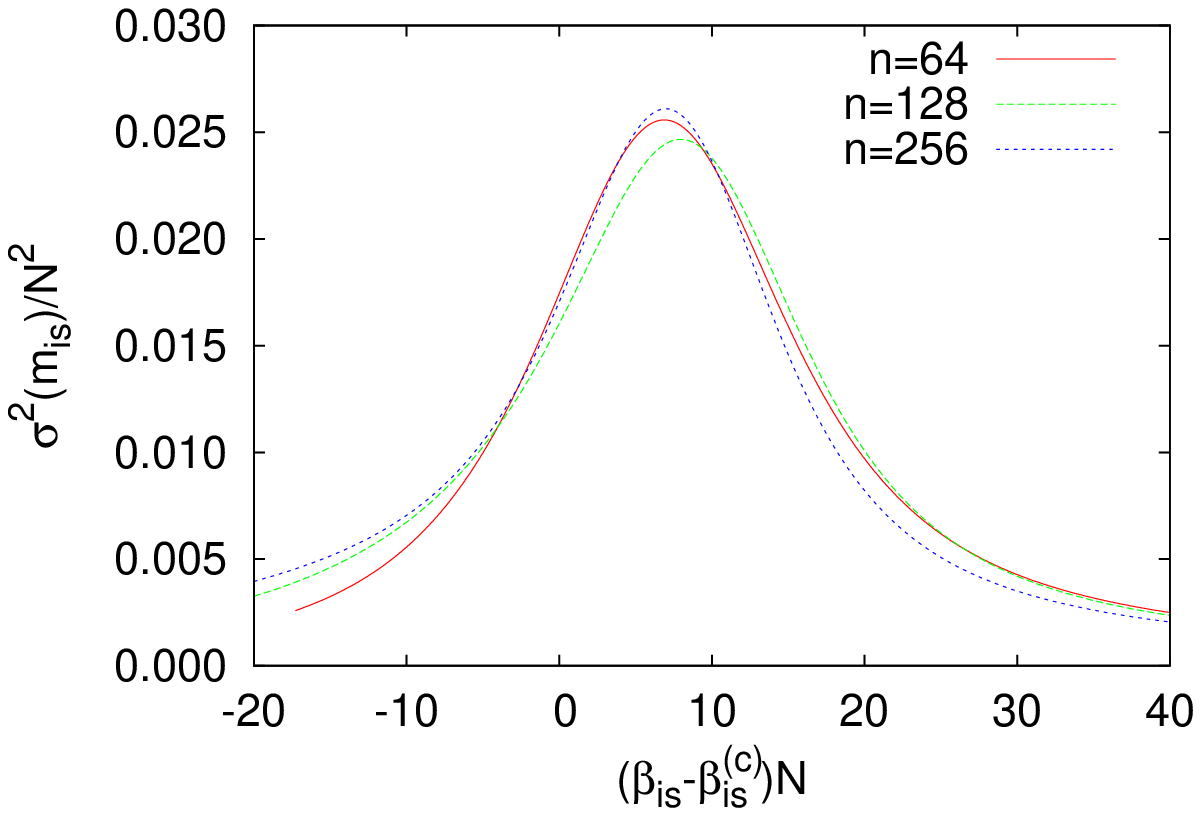}
  \caption{Plots of the fluctuations per monomer divided by $N$ in the number of ISAW
contacts ({\bf $m_{is}$}) at $\beta_{ss}=2.0$ for $2d$˛ with the horizontal axis scaled as (($\beta_{is}-\beta_{is}^{(c)})N$).
We have used $\beta_{is}^{(c)}=0.27$. Shown are lengths $64$, $128$ and   $256$.
}
\label{fig_2dtarns_bs2.0}
\end{figure}

Since the extended-globule $\theta$ transition has been well studied
we have focussed on the two other transitions: extended-crystal and
globule-crystal. Both these transitions are first order on the cubic
lattice \cite{bastolla:1997-01}. In the \textit{IHB--INH} model
\cite{krawczyk_2007:hb} the extended-crystal was also first order on
both the square and cubic lattice. So firstly let us consider this
transition in the semi-flexible \textit{ISAW} model on the square
lattice. By considering the scaling of the fluctuations in $m_{ss}$
and the distribution of $m_{ss}$ we confirm that the first order
nature of this transition. Figure~\ref{fig_2dtarns_bs2.0} demonstrates
that the data is consistent with first order scaling at the transition
point of the swollen to crystal phases.
%

\begin{figure}[ht!]
  \centering
  \includegraphics[scale=0.7]{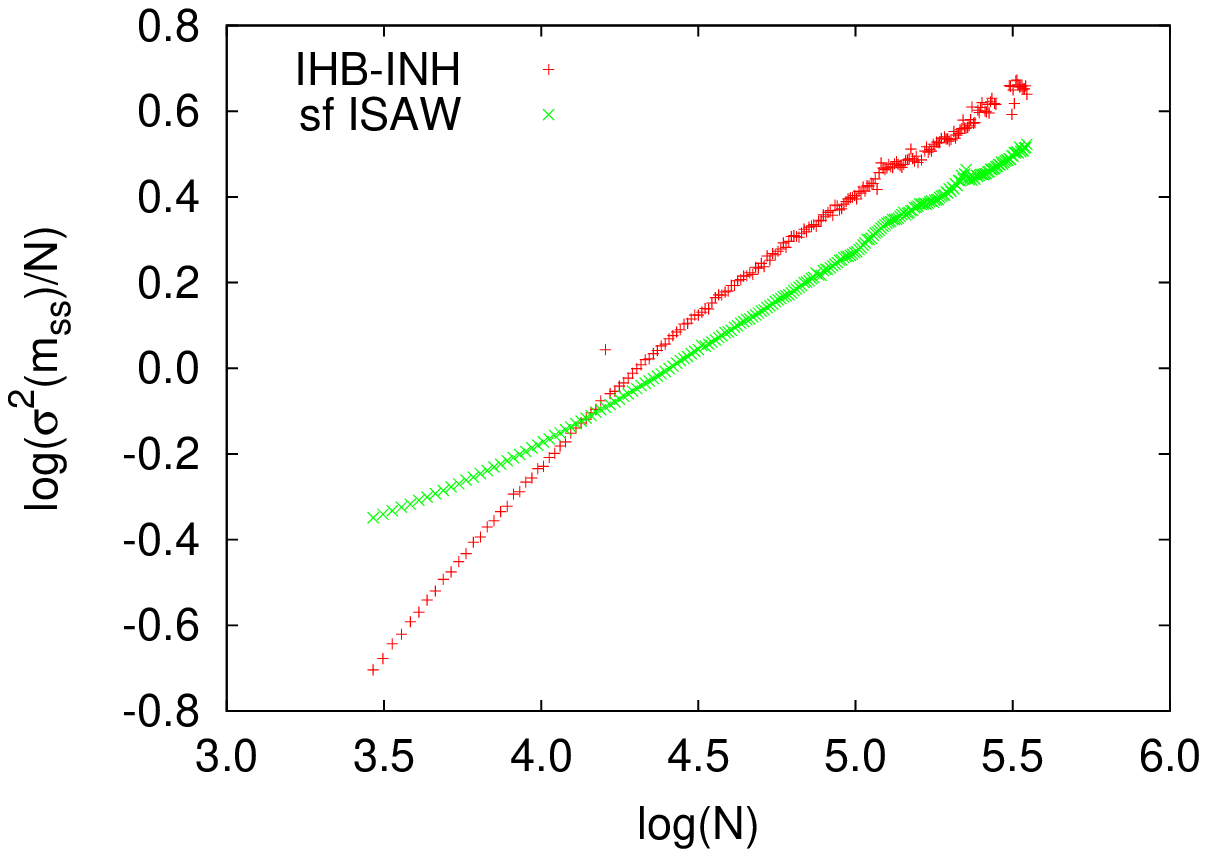}
  \caption{The maximum of fluctuations for the transition between the
    globule and crystalline phases for two models: semi-flexible
    \textit{ISAW} (a) and the \textit{IHB--INH} (b) models. (The first
    curve (a) is shifted to show them next to each other.)}
  \label{fig_flc_hb}
\end{figure}

In Figure~\ref{fig_flc_hb} the maximum of fluctuations for the
transition between the globule and crystalline phases for the
semi-flexible and \textit{IHB--INH} models are plotted together. The
semi-flexible data is taken from the one parameter runs with
$\beta_{is}=0.7$ at lengths up to $n=512$ and the \textit{IHB--INH}
data from simulations at $\beta_{nh}=1.0$ at lengths up to 256.  While
the data still has some corrections to scaling present the divergence
of the peak of the fluctuations seem to be controlled by the same
value of exponent. Ignoring corrections to scaling on the length range
128 to 256 would give us an estimate of the exponent controlling this
divergence for the \textit{IHB--INH} and semi-flexible ISAW models
respectively, of $0.41(2)$ and $0.43(2)$.

However, by being conservative with taking into account the corrections to scaling
for the largest lengths we have estimated the
exponent to be $0.4(2)$. Now, this exponent is expected to be $\alpha\phi$
where $\alpha$ is the specific heat exponent of the thermodynamic
limit transition and $\phi$ is the crossover exponent. Additionally it
is usually expected that $2-\alpha=1/\phi$ so we have
$\alpha=0.6(2)$ and $\phi=0.7(1)$.

\section{Semi-flexible \textit{INH} model ($\beta_{hb}=0$)}
Given that we have investigated the effect of stiffness on the
\textit{IHB} model and previously \cite{krawczyk_2007:hb} considered
the \textit{IHB--INH} model \cite{krawczyk_2007:hb} it was desirable
to consider the effect of stiffness on the \textit{INH} model for
completeness.

\begin{figure}[ht!]
  \centering
  \includegraphics[scale=0.7]{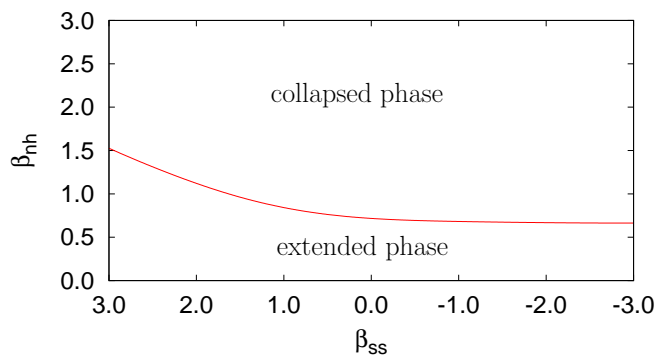}
    \includegraphics[scale=0.3]{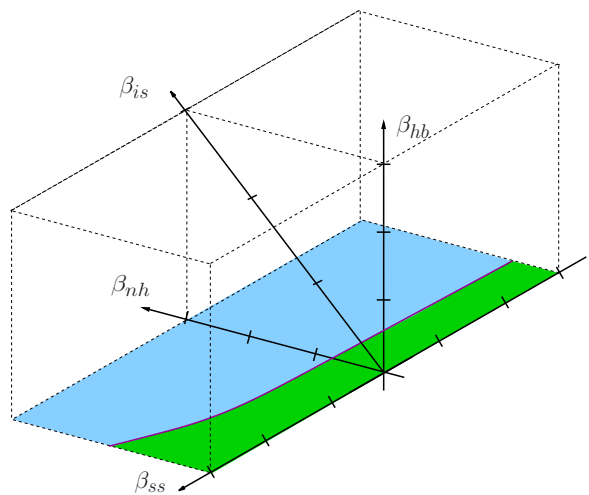}
  \caption{Plot of the finite size phase boundary for the
    semi-flexible \textit{INH} model on the square lattice (top). The
    plane of the parameters of the semi-flexible \textit{INH} model in
    the more general three parameter space: the two phases are denoted
    by a light (blue) shading for the globular phase and a
    mid-density(green)  shading for the extended swollen polymer
    phase. (bottom)}  
  \label{fig_mod_ssnh}
\end{figure}

When $\beta_{ss}=0$ the \textit{INH} model behaves exactly as the
\textit{ISAW} with a single collapse transition from the extended
phase at high temperatures to the globular collapsed phase at low
temperatures. While it is more difficult to detect the $\theta$ point
in two dimensions since the specific heat exponent $\alpha < 0$ it
signature can still be seen in the specific heat data. We have
determined a finite size phase boundary
$\beta_{nh}^{(c)}(\beta_{ss};n=128)$ that can be seen in
Figure~\ref{fig_mod_ssnh}. We immediately see that while the
\textit{INH} model behaves similarly to the \textit{ISAW} model the
addition of stiffness does not produce the same effect. At most one
phase transition is found for any positive or negative stiffness
energy.

On both lattices we therefore have found that the addition of negative
stiffness (effectively encouraging bends) to the interacting
non-hydrogen-bond model leaves the single phase transition from a
swollen phase at high temperatures to a globular phase at low
temperatures unchanged. This can be simply understood by noting that
most of the non-hydrogen-bond contacts occur between bends in the walk
anyway. The transition temperature goes to zero as the stiffness
energy goes to negative infinity.

When favouring straight segments, so that $\varepsilon_{ss}$ and
$\beta_{ss}$ are positive, then when any transition occurs it is again
of a similar type as the fully-flexible model (that is
$\theta$-point-like). However, at least for $\lambda=
\varepsilon_{ss}/\varepsilon_{nh}$ large enough no phase transition
occurs on lowering the temperature. 

A similar argument to the one in
Section 3 seems to hold: non-hydrogen-bond contacts do not occur between
stiffness sites which are favoured by large positive values of
$\varepsilon_{ss}$. We note that the zero temperature states are
pathological and there exits zero temperature phases transitions
(rod-coil and rod-globule). Nevertheless it seems that the change in
the zero-temperature state on varying the parameters still accords
with the position of the finite temperature swollen-globule phase
boundary. At zero temperatures the ground state for $\lambda>2$ is a
walk consisting only of straight segments with energy
$-n\varepsilon_{ss}$. For $0<\lambda<2$ the ground state consists of
long zig-zag paths with each `zig' and each `zag' being made up of two
steps (in this way each straight segment is always adjacent to two
bends) next to each other which have one non-hydrogen bond per step
and one stiffness parameter per two steps: the energy of this state is
$-n\varepsilon_{hb}-\frac{n}{2}\varepsilon_{ss}$. These two states
cross energies at $\lambda=2$: this seems to explain the asymptote of the
phase boundary in Figure~\ref{fig_mod_ssnh} for large positive
$\varepsilon_{ss}$.

\section{Conclusions}

We have investigated the effect of increasing stiffness and also
enhancing bends  on the lattice model of hydrogen-bonded polymers. We
have found that in both cases if there is a phase transition it is
unchanged from the fully-flexible model: namely a first order phase
transitions occurs in both two and three dimensions. We also argue
that if bending is sufficiently enhanced no phase transition occurs at
all. This is in contrast to the effect of adding stiffness to the
canonical model of self-interacting polymers where adding stiffness
results in three phases: a high temperature excluded-volume dominated
``swollen" phase, a liquid-like globule phase and an anisotropic
solid-like polymer crystal phase. We have investigated this
semi-flexible \textit{ISAW} problem in two dimensions and shown that
these three phases exist as they had previously been shown to exist in
three dimensions. We have investigated the globule-crystal transition
on the square lattice more closely and found that unlike three
dimensions where it is first order but like another recently studied
model,  extending the hydrogen bonding model by the addition of
non-hydrogen bond interactions, the transition is second order with
specific heat exponent $\alpha=0.6(2)$. 

\begin{figure}[ht!]
  \centering
\includegraphics[scale=0.5]{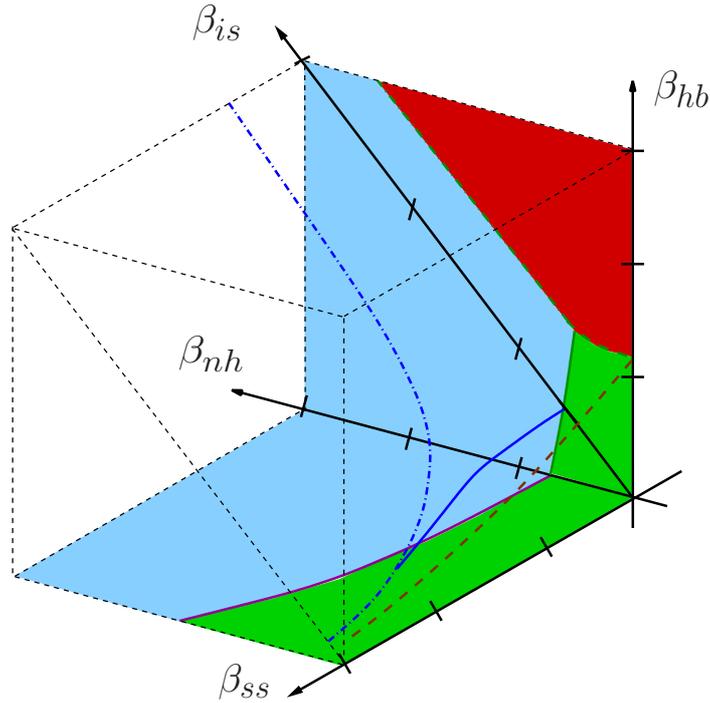}  
  \caption{The phase boundaries  found in the full three-dimensional
    phase space. We conjecture that only three phases occur in this
    larger space. we have illustrated the phases in the planes
    $\beta_{hb}=0$ and $\beta_{ss}=0$ in the quadrant where all the
    energies are positive:  the three phases are denoted by a light
    (blue) shading for the globular phase, a mid-density (green)
    shading for the extended swollen polymer phase and a dark (red)
    shading for the crystalline phase.}  
  \label{fig_mod_c}
\end{figure}

Putting together all the information at hand it is likely that in the
three-dimensional phase space of hydrogen-bond, non-hydrogen-bond
nearest neighbour interactions and stiffness only the three phases
already studied occur. In Figure~\ref{fig_mod_c} all the phase
boundaries found when the energies are all positive are illustrated.
One can the infer that for $\beta_{hb}$ and $\beta_{nh}$ small no
matter what the value of $\beta_{ss}$ the extended phase exists. Also,
one can infer that for large $\beta_{nh}$ the globular phase exists
and for large $\beta_{nh}$ the crystal phase exists. In this way the
partial results in the literature can now be understood.

\section*{Acknowledgements} 
Financial support from the Australian Research Council via its support
for the Centre of Excellence for Mathematics and Statistics of Complex
Systems is gratefully acknowledged by the authors.

\end{document}